\documentclass[prl,twocolumn,superscriptaddress]{revtex4-1}

\usepackage{graphicx}
\usepackage{float}
\usepackage{subfigure}
\usepackage{dcolumn}
\usepackage{gensymb}
\usepackage{bm}
\usepackage{epstopdf}
\usepackage{amsmath}
\usepackage{amssymb}
\usepackage{color} %,soul}

\usepackage[usenames,dvipsnames,svgnames,table]{xcolor}
\usepackage{tikz}
\usetikzlibrary{shapes}

\usepackage{bibunits}
\defaultbibliography{references.bib}
\defaultbibliographystyle{apsrev4-1}

\usepackage[margin=1.0in]{geometry}
\usepackage{setspace}

\usepackage{hyperref}
\hypersetup{
     colorlinks   = true,
     citecolor    = blue,
     linkcolor    = blue
}

\newcommand{\editor}[2]{%
  \expandafter\newcommand\csname #1note\endcsname[1]{%
    \textcolor{#2}{(\textbf{#1:} ##1)}}%
  \expandafter\newcommand\csname #1\endcsname[1]{%
    \textcolor{#2}{##1}}%
  \expandafter\newcommand\csname #1cancel\endcsname[1]{%
    \textcolor{#2}{\sout{##1}}}%
  \expandafter\newcommand\csname #1change\endcsname[2]{%
    \textcolor{#2}{\sout{##1} ##2}}%
  \newenvironment{#1text}{\color{#2}}{\color{black}}
}

\definecolor{tangerine}{rgb}{0.944,0.522,0}
\definecolor{verde}{rgb}{0.,0.6,0}
\definecolor{purple}{rgb}{.82, .13, .82}
\definecolor{turchese}{rgb}{0,0.5,0.5}

\definecolor{color1}{rgb}{0.827,0.89,0.592}
\newcommand{\markerone}{\raisebox{0.5pt}{\tikz{\node[draw,scale=0.3,regular polygon, regular polygon sides=3,fill=color1,rotate=0](){};}}}

\definecolor{color2}{rgb}{0.553,0.804,0.757}
\newcommand{\markertwo}{\raisebox{0.5pt}{\tikz{\node[draw,scale=0.3,regular polygon, regular polygon sides=3,fill=color2,rotate=90](){};}}}

\definecolor{color3}{rgb}{0.922,0.431,0.267}
\newcommand{\markerthree}{\raisebox{0pt}{\tikz{\node[draw,scale=0.4,diamond,fill=color3](){};}}}

\editor{SB}{tangerine}
\editor{LI}{blue}
\editor{DD}{verde}
\editor{GB}{purple}
\editor{RF}{turchese}

\makeatletter
\newcommand*{\addSI}{%
  \close@column@grid
  \cleardoublepage
  \twocolumngrid
}
\makeatother

\begin{document}

\begin{bibunit}

%\title{Generalized lattice dynamics approach to heat transport in crystals and glasses}
%\title{Generalized quasi harmonic approach to heat transport in crystals and glasses}
%\title{Bridging the gap between theoretical approaches to calculate heat transport in crystals and glasses}
% \title{Bridging the theory gap between heat transport \\ simulations in crystals and glasses \SB{non va bene? il concetto o la lingua?}\DD{\`E ok, anche se il termine "simulations" potrebbe essere sostituito da "calculations" o "modeling". }}
\title{Modeling heat transport in crystals and glasses \\ from a unified lattice-dynamical approach}

\author{Leyla Isaeva}
\affiliation{SISSA -- Scuola  Internazionale  Superiore  di  Studi  Avanzati,  Trieste, Italy}
\author{Giuseppe Barbalinardo}
\affiliation{Department of Chemistry, University of California at Davis, USA}
\author{Davide Donadio}
\affiliation{Department of Chemistry, University of California at Davis, USA}
\author{Stefano Baroni}
\affiliation{SISSA -- Scuola  Internazionale  Superiore  di  Studi  Avanzati,  Trieste, Italy}
\affiliation{CNR-IOM DEMOCRITOS, SISSA, Trieste, Italy}

\begin{abstract}
    We introduce a novel approach to model heat transport in solids, based on the Green-Kubo theory of linear response. It naturally bridges the Boltzmann kinetic approach in crystals and the Allen-Feldman model in glasses, leveraging interatomic force constants and normal-mode linewidths computed at mechanical equilibrium. At variance with molecular dynamics, our approach naturally and easily accounts for quantum mechanical effects in energy transport. Our methodology is carefully validated against results for crystalline and amorphous silicon from equilibrium molecular dynamics and, in the former case, from the Boltzmann transport equation.
\end{abstract}

\date{\today }

\maketitle

Heat transport in solid insulators, either crystalline or disordered, is dominated by the dynamics of lattice vibrations. Far from melting, atomic displacements from equilibrium are much smaller than interatomic distances and they can thus be treated in the (quasi-) harmonic approximation. In crystals this observation enables a kinetic description of heat transport in terms of phonons that can be embodied in the Peierls--Boltzmann transport equation (BTE) \cite{*[{See for instance Chapter VII of }][]Ziman1960,Fugallo2018}. In disordered systems the typical phonon mean free paths may be so short that the quasi-particle picture of heat carriers breaks down and  BTE is no longer applicable, making it necessary to resort to molecular dynamics (MD),  in either its nonequilibrium or equilibrium (EMD) flavors \cite{HMM,Fugallo2018}. MD is of general applicability to solids, either periodic or disordered, and liquids. Yet, it may require long simulation times and it is subject to statistical errors, which are at times cumbersome to evaluate especially for systems at low temperatures, where lack of ergodicity may be an issue. Most importantly, MD cannot account for quantum-mechanical effects \cite{BedoyaMartinez:2014cq}, which are instead easily treated in BTE, thus making the treatment of heat transport for glasses in the quantum regime, \emph{i.e.} below the Debye temperature, a methodological challenge.

In this paper we present a novel approach to heat transport in insulating solids, which combines the Green-Kubo (GK) theory of linear response \cite{Green1952,*Green1954,*Kubo1957a,*Kubo1957b,HMM} and a quasi-harmonic description of lattice vibrations, thus resulting in a compact expression for the thermal conductivity that unifies the BTE approach in the single-mode relaxation-time approximation (RTA) for crystals \cite{Fugallo2018} and a generalization of the Allen-Feldman (AF) model for disordered system \cite{Allen1989,*Allen1993} that explicitly and naturally accounts for normal-mode lifetimes. The main ingredients of our approach are the matrix of the inter-atomic force constants (IFC) computed at mechanical equilibrium and the anharmonic lifetimes of the vibrational modes, as computed from the cubic corrections to the harmonic IFCs using the Fermi's golden rule \cite{Fabian1996}. Our theory is thoroughly validated in crystalline and amorphous silicon by comparing its predictions with those of EMD simulations and BTE computations.

The basis of our work is the GK theory of linear response \cite{Kubo1957a,*Kubo1957b,HMM}, which relates the heat conductivity to the ensemble average of the heat-flux auto-correlation function:
\begin{equation}
   \kappa_{\alpha\beta} = \frac{1}{V k_B T^2}  \int\limits_{0}^{+\infty} \langle J_\alpha(t) J_\beta(0)\rangle dt, \label{eq:GK}
\end{equation}
where $V$ and $T$ are the system volume and temperature, $k_B$ is the Boltzmann constant, $J_\alpha(t)$ the $\alpha$-th Cartesian component of the macroscopic heat flux, which in solids coincides with the energy flux, and $\langle\cdot\rangle$ indicates a canonical average over initial conditions \cite{HMM}. Far from melting,  the energy flux and the lattice Hamiltonian of a solid, both crystalline and amorphous, can be expressed as power series in the atomic displacements, and Eq. \eqref{eq:GK} can be evaluated in terms of Gaussian integrals using standard field-theoretical techniques.

The energy flux can be expressed in terms of atomic positions, $\mathbf{R}_i$, and local energies, $\epsilon_i$, as $\mathbf{J}=\sum_i ( \dot{\mathbf{R}}_i \epsilon_i+\mathbf{R}_i\dot\epsilon_i )$ \cite{HMM}, where in the harmonic approximation $\epsilon_i$ can be defined as: $ \epsilon_i = \frac{M_i}{2}  \sum_{\alpha}  \left ( \dot{u}_{i\alpha} \right )^2 + \frac{1}{2} \sum_{j\alpha\beta}   u_{i\alpha} \Phi_{i\alpha}^{j\beta}u_{j\beta}$, $M_i$ being the mass of $i$-th atom, $\mathbf u_i=\mathbf R_i-\mathbf R^\circ_i$ its displacement from its equilibrium position, $ \mathbf R^\circ_i $, $\alpha$ and $\beta$ label Cartesian components, and $\Phi_{i\alpha}^{j\beta} =  \left . \frac{\partial^2 E}{\partial u_{i\alpha}\partial u_{j\beta}} \right |_{\mathbf u=0}$ is the IFC matrix. By expressing the energy flux in terms of the $\mathbf u$'s, one obtains: $\mathbf{J}= \mathbf J^\circ +\frac{d}{dt}\sum_i \mathbf u_i\epsilon_i$, where $\mathbf J^\circ = \sum_i \mathbf{R}^\circ_i\dot\epsilon_i$. The second term on the right-hand side of this expression is the total time derivative of a process that, in the absence of atomic diffusion, is stationary and of finite variance. A recently established \emph{gauge invariance} principle for heat transport \cite{Marcolongo2016,*Ercole2016} states that such a total time derivative does not contribute to the thermal conductivity. We will therefore dispose of it and express the energy flux as: $ \mathbf{J}\gets \mathbf J^\circ$. Note that it is essential to use equilibrium atomic positions in the definition of $\mathbf J^\circ$, \emph{i.e.} the positions describing the (metastable) mechanical equilibrium of any given model of an ordered or disordered system, rather than instantaneous ones. Otherwise, the difference $\mathbf J-\mathbf J^\circ$ would not be a total time derivative and the value of the transport coefficient resulting from $\mathbf J^\circ$ would be biased. By making use of Newton's equation of motion, the final expression for the harmonic heat flux reads \cite{Allen1989}:
\begin{equation}
    J_{\alpha} =  \frac{1}{2} \sum_{ij\beta\gamma} (R^{\circ}_{i\alpha} - R^{\circ}_{j\alpha})\Phi_{i\beta}^{j\gamma} {u}_{i\beta} \dot{u}_{j\gamma}, \label{eq:heatflux}
\end{equation}
where the minimum-image convention is adopted for computing inter-atomic distances in periodic boundary conditions.

By inserting Eq.~\eqref{eq:heatflux} into Eq.~\eqref{eq:GK}, the integrand is cast into the canonical average of a quartic polynomial in the atomic positions and velocities. In the harmonic approximation, this average reduces to a Gaussian integral, which can be evaluated by way of the Wick's theorem \cite{*[{See for instance Sec. 2.3 in }] [{}] Negele1988}. By doing so, the resulting time integral would diverge, thus yielding an infinite conductivity, as expected for a harmonic crystal \cite{Rieder:1967fi}. In order to regularize this integral, we introduce anharmonic effects by renormalizing the single-mode correlation functions using the normal-mode lifetimes, as explained below. Our final result for the heat conductivity tensor reads:

\begin{gather}
    \kappa_{\alpha\beta} = \frac{k_B}{V} \sum\limits_{nm} {v}^\alpha_{nm} {v}^\beta_{nm} {\tau}^\circ_{nm}, \label{eq:classical} \\
    v^\alpha_{nm} = \frac{1}{2 \sqrt{\omega_{n} \omega_{m} }} \sum_{ij\beta\gamma} \frac{ R^{\circ}_{i\alpha} - R^{\circ}_{j\alpha}}{\sqrt{M_i M_j}} \Phi_{i\beta}^{j\gamma} e_n^{i\beta} e_m^{j\gamma}, \label{eq:vnm} \\
    \tau^\circ_{nm} =  \frac{\gamma_n+\gamma_m}{(\gamma_n+\gamma_m)^2 + (\omega_n-\omega_m)^2 } +\mathcal{O}(\epsilon^2),  \label{eq:taunm}
\end{gather}
where $\omega_n$ and $\gamma_n$ are the harmonic frequency and decay rate of the $n$-th normal mode, and $e_{ni}^\alpha$ is the corresponding eigenvector of the dynamical matrix, $\sum_{j\beta} \frac{1}{\sqrt{M_i M_j}} \Phi^{i\alpha}_{j\beta} e_{nj}^\beta = \omega_n^2 e_{ni}^\alpha $, and $\epsilon$ indicates the ratio $\gamma/\omega$, which vanishes in the harmonic limit. Eqs. (\ref{eq:classical}-\ref{eq:taunm}) will be dubbed as the {\sl quasi-harmonic Green-Kubo} (QHGK) approximation for the heat conductivity.

In order to establish Eq. \eqref{eq:classical}, we first express the energy flux in Eq. \eqref{eq:heatflux} in terms of normal-mode coordinates and momenta, defined as: $\xi_n =\sum_{i\alpha} \sqrt{M_i}u^i_\alpha e_{ni}^\alpha$ and $\pi_n =\sum_{i\alpha}\dot u_i^\alpha e_{ni}^\alpha / \sqrt{M_i} $, reading: $ J^\alpha = \sum_{nm} v_{nm}^\alpha \sqrt{\omega_n \omega_m} \xi_n \pi_m$. It is then convenient to express these normal-mode coordinates and momenta in terms of classical complex amplitudes, reminiscent of the quantum boson ladder operators and defined as: $\alpha_n = \sqrt{\frac{\omega_n}{2}} \xi_n + \frac{i}{\sqrt{2\omega_n}}\pi_n$, whose time evolution is $\alpha_n(t) = \alpha_n(0) \mathrm{e}^{-i\omega_nt}$. By doing so, the energy flux can be expressed in terms of the $\alpha$ amplitudes as
\begin{equation}
    J_\beta=\frac{i}{2} \sum_{nm}v^\beta_{nm}\omega_m (\alpha^*_n+\alpha_n)(\alpha^*_m-\alpha_m). \label{eq:Jalpha}
\end{equation}
By using this expression, the integrand in Eq. \eqref{eq:GK} becomes a Gaussian integral of a fourth-order polynomial in the $\alpha$'s and $\alpha^*$'s that, by means of the Wick's theorem \cite{*[{See for instance Sec. 2.3 in }] [{}] Negele1988}, can be cast into a sum of products of pairs of single-mode (classical) Green's functions, $\langle\alpha^*_n (t) \alpha_m (0) \rangle = \delta_{nm} g_n(t)$ and $\langle\alpha_n (t) \alpha_m (0) \rangle = 0$. In the purely harmonic approximation, one would have $g^\circ_n(t) = \frac{k_BT}{\omega_n} \mathrm{e}^{i\omega_nt}$. Anharmonic effects broaden the vibrational lines by a finite line-width, $\gamma_n$, which results in a finite imaginary part of the frequency and in a decay of the single-mode Green's function, reading: $g_n(t) = \frac{k_BT}{\omega_n} \mathrm{e}^{i(\omega_n+i\gamma_n)t}$. By plugging this expressions into the lengthy formula that results from applying Wick's theorem to the integrand of Eq. \eqref{eq:GK} and performing the time integral, after some cumbersome but straightforward algebra we get Eq. \eqref{eq:classical}. A full derivation of Eqs. (\ref{eq:classical}-\ref{eq:taunm}) is presented in the Supplemental Material (SM), Sec. S1.

To lowest order in the anharmonic interactions, vibrational linewidths can be computed from the classical limit of the Fermi golden rule, $ \gamma_n =  \frac{\pi \hbar^2}{8 \omega_n} \sum_{ml} \frac{|V'''_{nml}|^2}{\omega_m \omega_l} \bigl [ \frac{1}{2} (1 + n_m + n_l) \delta(\omega_n - \omega_m - \omega_l) + (n_m - n_l) \delta(\omega_n + \omega_m - \omega_l)  \bigr ]$, where $n_l$ is the Bose-Einstein occupation number of the $l$-th normal mode and $V'''_{nlm}=\frac{\partial^3 V}{\partial \xi^n\partial \xi^l\partial \xi^m}$ is the third derivative of the potential energy with respect to the amplitude of the lattice distortion along the lattice normal modes \cite{Fabian1996}.

In order to show that our QHGK expression for the thermal conductivity, Eq. \eqref{eq:classical}, reduces to the predictions of the BTE-RTA in crystals, we first notice that the $v^\alpha$ matrices of Eq. \eqref{eq:vnm} can be expressed in terms of the matrix elements of the matrices $\bigl (V^\alpha \bigr )_{i\gamma}^{j\delta} = \frac{R^\circ_{i\alpha}-R^\circ_{j\alpha}}{2\sqrt{M_iM_j}}\Phi_{i\gamma}^{j\delta}$ between normal-mode eigenvectors: $v^\alpha_{nm} = \bigl ( e_n,V^\alpha\cdot e_m \bigr )/\sqrt{\omega_n\omega_m}$, where the notations ``$(e,e')$'' and ``$V\cdot e$'' indicate scalar and matrix-vector products in the space of atomic displacements. In crystals equilibrium atomic positions are characterised by a discrete lattice position, $\mathbf{a}_i$, and by an integer label, $s_i$, indicating different atomic sites within a unit cell, $\mathbf{d}_s$: $\mathbf{R}^\circ_i = \mathbf{a}_i + \mathbf{d}_{s_i}$. Likewise,  in the Bloch representation, normal modes can be labelled by a quasi-discrete wavevector, $\mathbf q$, belonging to the first Brillouin zone (BZ), and by a band index, $\nu$: $n\to (\mathbf{q}_n,\nu_n)$. In particular, the IFC matrix and its eigenvectors can be expressed as $\frac{1}{\sqrt{M_iM_j}}\Phi_{j\beta}^{i\alpha} = \sum_\mathbf{q} \mathrm{e}^{i\mathbf{q}\cdot (\mathbf{R}^\circ_i-\mathbf{R}^\circ_j)} D_{t\beta}^{s\alpha}(\mathbf{q})$, where $D_{t\beta}^{s\alpha}(\mathbf{q})$ is the dynamical matrix of the system and $\eta^{s\alpha}_{\nu\mathbf{q}}$ its eigenvectors: $e^\alpha_{nu}=\mathrm{e}^{i\mathbf{q}_n\cdot \mathbf{R}^\circ_i} \eta^{s_i\alpha}_{\nu_n\mathbf{q}_n} $ and $\sum_{t\beta} D^{s\alpha}_{t\beta}(\mathbf{q}) \eta_{\nu\mathbf{q}}^{t\beta} = \omega_{\nu\mathbf{q}}^2 \eta^{s\alpha}_{\nu\mathbf{q}} $. When normal-mode eigenvectors are chosen to be real, the $v^\alpha$ matrices of Eq. \eqref{eq:vnm} are real and anti-symmetric. In particular, $v^\alpha_{nn}=0$ and a non-vanishing thermal conductivity results from the matrix elements of $v^\alpha$ connecting (quasi-) degenerate normal modes, \emph{i.e.} modes %that are degenerate
with frequencies that coincide within the sum of their line widths.
%inverse lifetimes).
In the Bloch representation, $v^\alpha$ is anti-Hermitian and block-diagonal with respect to the wave-vector, $\mathbf q$. Its diagonal elements are imaginary, though not necessarily vanishing. In this representation one has: $v^\alpha_{\nu\mathbf{q},\mu\mathbf{p}} = i\frac{\delta_{\mathbf{q}\mathbf{p}} }{\sqrt{\omega_{\nu\mathbf{q}}\omega_{\mu \mathbf{p}}}} \bigl ( \eta_{\nu\mathbf{q}}, D^\alpha(\mathbf{q})\cdot \eta_{\mu\mathbf{q}} \bigr ) $, where $D^\alpha(\mathbf{q})=\frac{\partial D(\mathbf{q})}{\partial q^\alpha}$.
At fixed $\mathbf q$, the vibrational spectrum is strictly discrete \emph{i.e.} it remains so even in the thermodynamic limit. The number of $\mathbf q$ points for which there exists a pair of distinct modes, $(\mathbf q,\nu)$ and $(\mathbf q,\mu)$ with $\nu\ne\mu$, that are degenerate within the sum of their line-widths $(|\omega_{\mathbf q\nu} - \omega_{\mathbf q\mu}| \lesssim  \gamma_{\mathbf q\nu} + \gamma_{\mathbf q\mu})$ is vanishingly small, because,  in practice, this quasi-degeneracy can only occur close to high-symmetry lines. Furthemore, for such few pairs, one can prove that $v_{\nu\mathbf q,\mu\mathbf q}\propto \omega_{\nu\mathbf q} - \omega_{\mu\mathbf q}$.
Hence in the periodic case the $\tau^\circ$ matrix in Eq.~\ref{eq:taunm} is strictly diagonal, $\tau^\circ_{\mathbf q\nu, \mathbf p\mu}=\delta_{\mathbf{q} \mathbf{p}} \delta_{\nu\mu}\tau^\circ_{\mathbf q \nu}$,  where $\tau^\circ_{\mathbf q\nu} =\frac{1}{2\gamma_{\mathbf q\nu}}$ is the anharmonic lifetime of the $(\mathbf q,\nu)$ normal mode.
We conclude that, for periodic systems in the Bloch representation, the double sum in Eq. \eqref{eq:classical} can be cast into a single sum over diagonal terms, reading: $\kappa_{\alpha\beta}=\sum_{\mathbf{q}\nu} v_{\nu\mathbf{q}}^\alpha v_{\nu\mathbf{q}}^\beta \tau_{\nu\mathbf{q}}$, where $v_{\nu\mathbf{q}}^\alpha = \frac{1}{2\omega_{\nu\mathbf q} }  \bigl ( \eta_{\nu\mathbf{q}}, D^\alpha(\mathbf{q})\cdot \eta_{\nu\mathbf{q}} \bigr )=  \frac{\partial \omega_{\nu\mathbf{q}}}{\partial q^\alpha}$ is the group velocity of the $\nu$-th phonon branch.
% and $\tau_{\nu\mathbf{q}} = \tau^\circ_{\nu\mathbf{q},\nu\mathbf{q}} = \frac{1}{ 2 \gamma_{\nu\mathbf{q}} }$ is the $\mathbf{q}\nu$-th phonon lifetime.
This is the final expression for the thermal conductivity of a crystal in  QHGK, which remarkably coincides with the solution of BTE-RTA.\cite{Ziman1960} We tested the QHGK approach against BTE-RTA for a crystalline silicon supercell of 1728 atoms, with a lattice parameter of 5.431 \AA. The two calculations, performed with different codes, give the same results, as expected by the proven equivalence of the two methods for crystalline systems (see Figure S1 in SM).

Moving to the quantum regime is straightforward in our approach. To this end, we start from the quantum GK formula \cite{Green1952,*Green1954,Kubo1957a,*Kubo1957b,HMM}, reading:
\begin{equation}
   \kappa_{\alpha\beta} = \frac{1}{V T} \int \limits_{0}^{1/k_BT} d \lambda \int \limits_{0}^{+\infty} dt \langle \hat{J}_\alpha(t+i\hbar\lambda) \hat{J}_\beta(0)\rangle ,
\label{eq:GKquantum}
\end{equation}
where $\hat{J}_\alpha$ are quantum heat-flux operators and $\langle\cdot\rangle$ indicates quantum canonical averages. A quantum expression for the heat flux is obtained from its classical counterpart, Eq. \eqref{eq:Jalpha}, by making the substitutions: $\alpha\to \sqrt{\hbar} \hat{a}$ and $\alpha^*\to \sqrt{\hbar} \hat{a}^\dagger$, $\hat{a}^\dagger$ and $\hat{a}$ being normal-mode creation/annihilation operators, satisfying the standard
commutation relations for bosons: $\left [ {\hat{a}}, {\hat{a}}^\dagger \right ] = 1$. Note that no ordering ambiguities arise when quantizing Eq. \eqref{eq:Jalpha} because the $v^\alpha_{nm}$ matrices are antisymmetric, and they therefore vanish for $n=m$. The resulting expression for the quantum heat flux is:
\begin{equation}
    \hat{J}_\beta = \frac{i\hbar}{2} \sum_{nm}v^\beta_{nm}\omega_m (\hat{a}^\dagger_n+\hat{a}_n)(\hat{a}^\dagger_m-\hat{a}_m). \label{eq:JQalpha}
\end{equation}
The computation of the heat conductivity proceeds exactly as in the classical case, except for the expressions for the single-mode Green's functions. In the quantum case they read: $\langle \hat{a}_n^\dagger(t) \hat{a}_n(0)\rangle=n_n \mathrm{e}^{i(\omega_n+i\gamma_n)t}$ and $\langle \hat{a}_n(t) \hat{a}^\dagger_n(0)\rangle=(n_n+1) \mathrm{e}^{-i(\omega_n-i\gamma_n)t}$, $n_n = 1 \left / \left (e^{\frac{\hbar \omega_n}{k_B T}} - 1 \right ) \right .$ being the Bose-Einstein distribution function.
The final quantum-mechanical expression for the heat conductivity in the QHGK is:
\begin{equation}
   \kappa_{\alpha \beta} = \frac{1}{V} \sum_{nm} c_{nm} v_{n m}^{\alpha} v_{nm}^{\beta} \tau^\circ_{n m}, \label{eq:quantum}
\end{equation}
with $ c_{n m} =  \frac{\hbar \omega_n \omega_m }{T} \frac{n_{m} - n_{n}}{\omega_{m}-\omega_{n} }$. For $n=m$ this term reduces to the modal heat capacity $c_n=k_B\left ( \frac{\hbar\omega_n}{k_B T} \right )^2 \frac{e^{\hbar\omega_n/k_BT}}{(\mathrm{e}^{\hbar\omega_n/k_BT}-1)^2}$. The other symbols are the same as in Eqs. (\ref{eq:vnm}-\ref{eq:taunm}) for the classical case. $\tau^\circ_{nm}$, in particular, is only different from zero for $|\omega_n-\omega_m| \lesssim \gamma_n+\gamma_m$. Following the same derivation as for the classical case, one can prove that for periodic crystals Eq. \eqref{eq:quantum} reduces to BTE-RTA. Furthermore, in the classical limit, one has $\lim_{T \to \infty} c_{nn} = k_B$ and the quantum formula, Eq. \eqref{eq:quantum}, reduces to Eq. \eqref{eq:classical}. Further details are given in Sec. S2 of SM.

\begin{figure}[!tp]
\includegraphics[width = 0.49\textwidth]{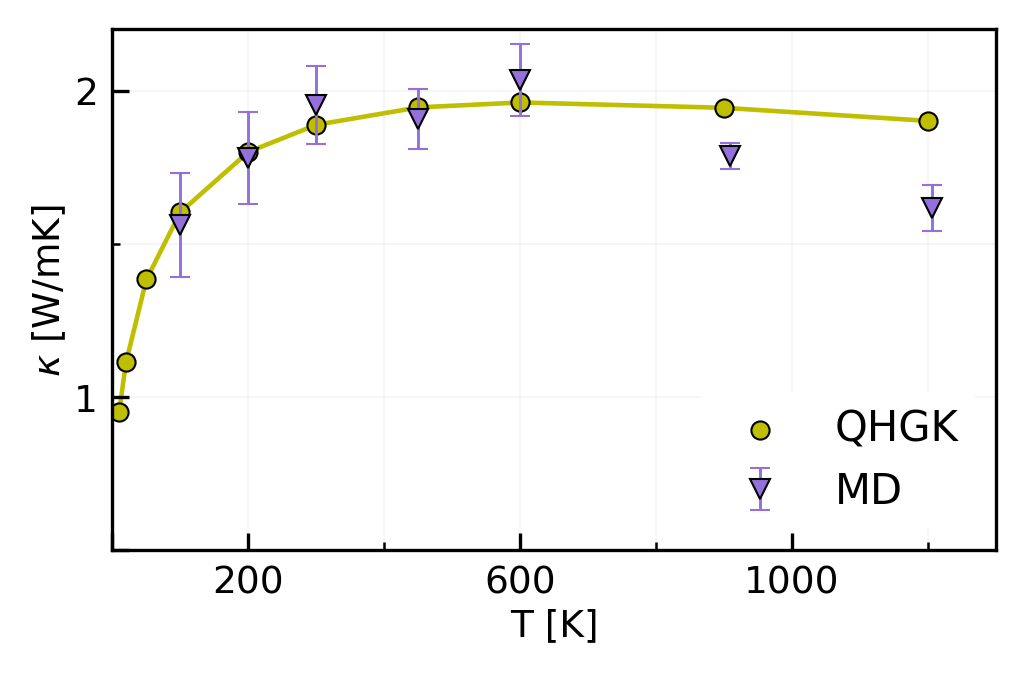}
\caption{Comparison between the thermal conductivity of a-Si computed for a 1728-atom supercell by the classical Green-Kubo theory of linear response using either our QHGK approach (Eq. \ref{eq:classical}, green) or equilibrium molecular dynamics (purple). The $k_{xx}$, $k_{yy}$ and $k_{zz}$ components of thermal conductivity tensor $\bm{\kappa}$ are averaged to obtain a value corresponding to an isotropic amorphous media.}
\label{fig:kappa_classical}
\end{figure}

We validate our QHGK approach by testing the results of Eqs. \eqref{eq:classical} and \eqref{eq:quantum} against  MD simulations for amorphous silicon. Interatomic interactions are modeled using the empirical bond-order Tersoff potential~\cite{Tersoff:1989tr}, which describes  well the thermal conductivity of bulk and nanostructured silicon, including a-Si \cite{Allen1993,He:2011wq,He:2012tq,Larkin2014}. We consider  a 1728-atom model of a-Si, generated by MD by quenching from the melt. Several EMD simulations where then run at different temperatures, as described in SM \cite{Fan:2015ba,Fan2017}. The integral of the heat flux autocorrelation function in Eq.~\eqref{eq:GK} can then be efficiently evaluated  via \emph{cepstral analysis}, as described in Refs. \onlinecite{HMM} and \onlinecite{Ercole2017}, which can be enhanced by averaging over multiple trajectories at low temperature ($T\le 300$ K).
%{\it via} either an average over multiple trajectories or via}
Details on the data analysis procedure followed here and on the estimate of the statistical errors is given in the S3 section of the SM. The results of these calculations are reported in Figure~\ref{fig:kappa_classical} and exhibit a weak non-monotonic dependence on $T$. Performing similar MD simulations on models of increasing size (4,096 and 13,824 atoms) generated with the same protocol, we have verified that size effects on $\kappa$ at 300 K amount to less than 15$\%$, which is of the same order as the variation $\kappa$ among different models with the same size. The computation of the IFC matrix, normal-mode frequencies and lifetimes is described in detail in SM, where we also display the resulting dependence of lifetimes on temperature (Figure \ref{fig:linewidths} of SM). The resulting strongly diagonally dominant form of the $\tau^\circ$ matrices in Eq. \eqref{eq:taunm} is also displayed in Figure \ref{fig:linewidths} of SM.

The thermal conductivity obtained by QHGK is in excellent agreement with that computed by EMD for $T\leq 600$ K (Figure~\ref{fig:kappa_classical}).
At higher temperatures  QHGK  overestimates $\kappa$, as it neglects higher-order anharmonic effects. We point out that, in spite of the formal analogies
with the AF model \cite{Allen1989,Allen1993,Feldman:1993tn} and recent refinements thereof \cite{Donadio:2010kp,Zhu:2016ks}, Eq. \eqref{eq:classical} 
entails no empirical parameters. It thus allows one to predict temperature trends dictated by anharmonic effects in good agreement with 
MD~
\cite{footnote2}
%\footnote{The temperature dependence in the AF model lies only in the heat capacity term, therefore in the classical limit $\kappa$ is temperature independent.}
, without making any \emph{a priori} distinction among propagating, diffusive, or localized vibrational modes. Similarly to the GK modal analysis approach \cite{Lv:2016dr}, based on classical MD, the transport character of the modes is dictated by the relative contribution from the diagonal and slightly off-diagonal terms of the $v^\alpha_{nm}$ matrix, weighted by $\tau^\circ_{nm}$ (Figure \ref{fig:linewidths}). %\ref{fig:heatmap_tau}).
The generality of QHGK is expected to have a major impact for the study of weakly disordered systems, which are beyond the scope of applicability of approaches based on the BTE and the AF model.

\begin{figure}[!tbp]
\includegraphics[width = 0.51\textwidth]{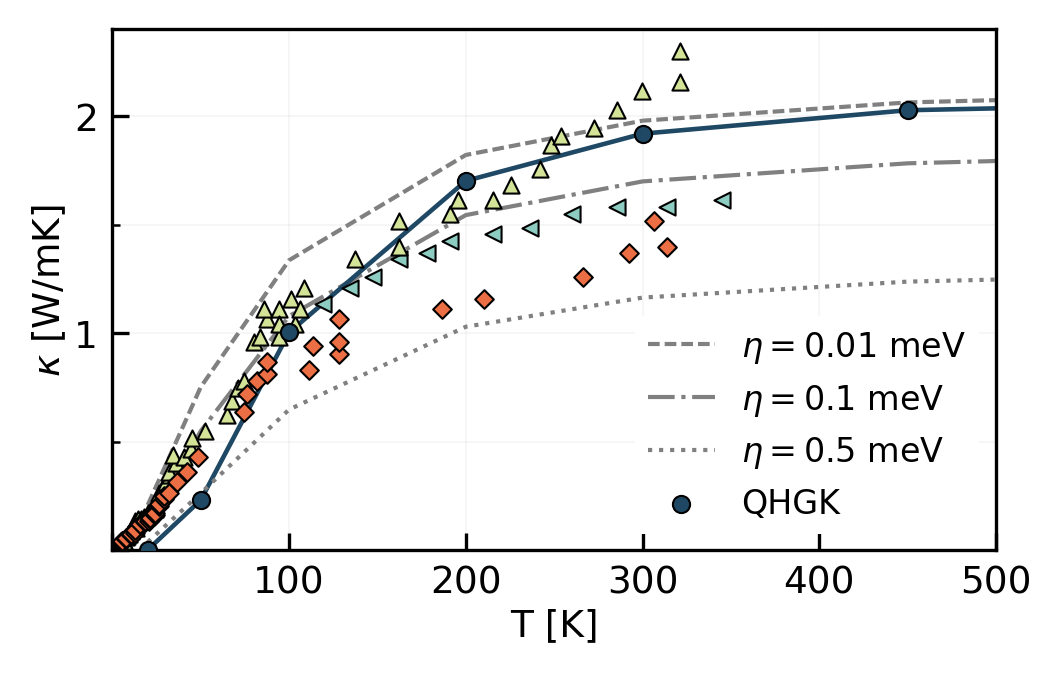}
\caption{Thermal conductivity computed for a 13824-atom supercell of a-Si using the quantum QHGK approach in the quantum regime (Eq. \ref{eq:quantum}), compared with the Allen-Feldman approach ~\cite{Allen1989,Allen1993,Feldman:1993tn} and experimental data
(\protect\markerone, \protect\markerthree ~Ref.\cite{Zink:2006wt}), (\protect\markertwo ~Ref.\cite{Cahill1994}). The broadening $\eta$ used in Allen-Feldman calculations is set equal for every normal mode.}
\label{fig:kappa_quantum}
\end{figure}

\begin{figure}[!tbp]
\includegraphics[width = 0.48\textwidth]{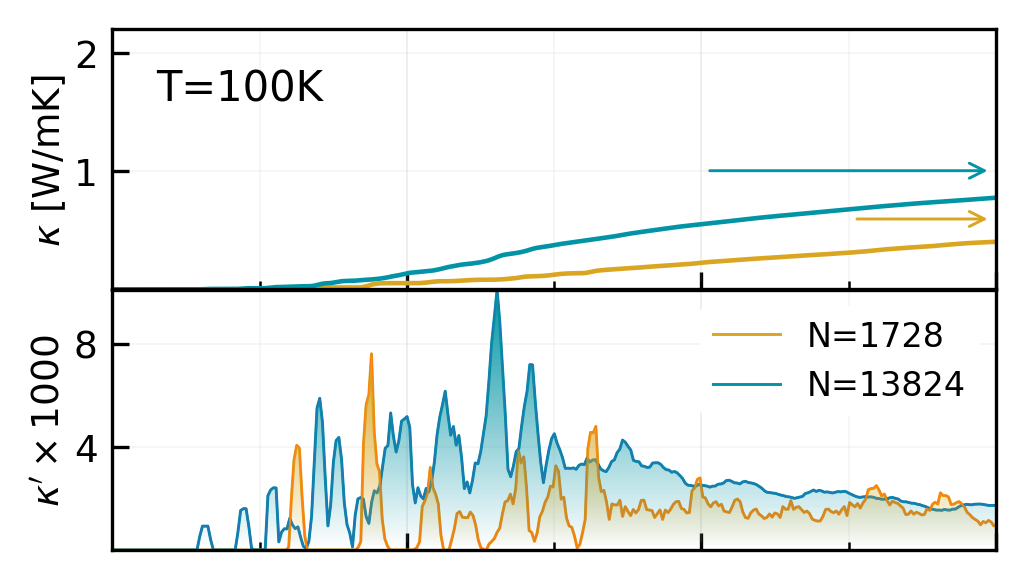}
\includegraphics[width = 0.485\textwidth]{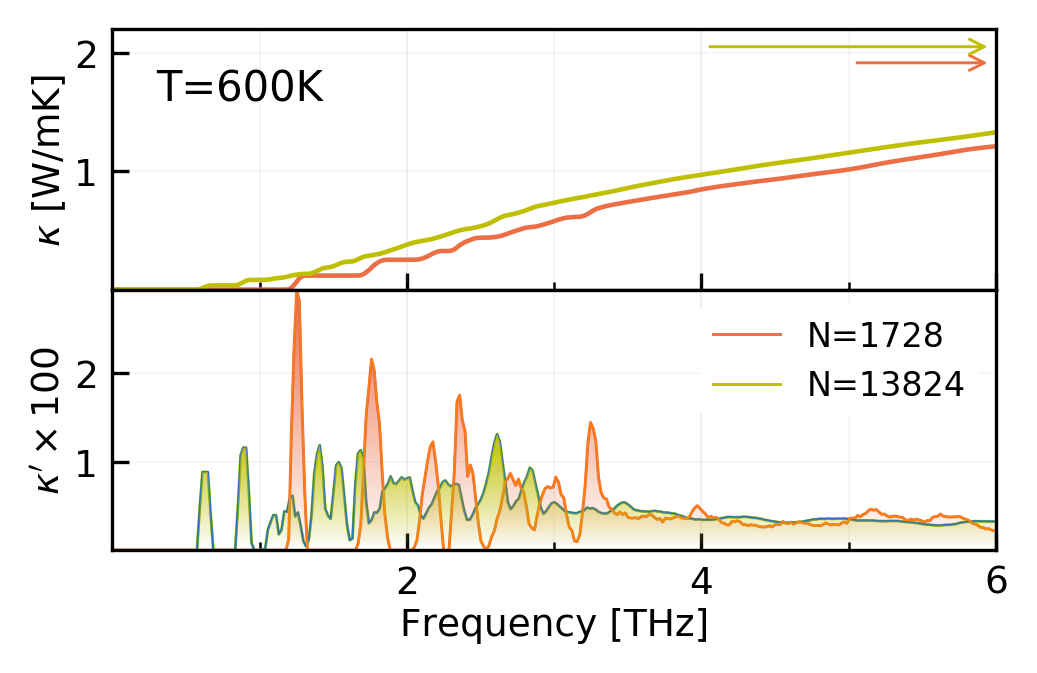}
\caption{%\GB{Maybe pelo nell'uovo, but shouldn't the unit of $\kappa'$ be different? Feel free to ignore this comment.}
Conductivity accumulation function, $\kappa(\omega)$, and frequency-resolved differential conductivity, $\kappa^{\prime}(\omega)$ (Eqs. \ref{eq:kappa'(omega)} and \ref{eq:kappa(omega)}), computed for two different model sizes ($N$=1,728 and $N$=13,824 atoms)
at temperatures $T$=100 K and $T$=600 K. Horizontal arrows in the upper panels indicate cumulative values of $\kappa$. $\kappa^\prime$ is in units of WK$^{-1}$m$^{-1}$ps. }
\label{fig:tmp}
\end{figure}

QHGK is a general theory that allows one to accurately calculate thermal transport in both crystals and glasses at a full quantum mechanical level.  In Figure \ref{fig:kappa_quantum} we report our results from quantum QHGK calculation for an amorphous Si model of 13824 atoms along with three sets of experimental data \cite{Cahill1994,Zink:2006wt}.
QHGK results are in excellent agreement with the measurements in \cite{Zink:2006wt} above 100 K. At lower temperature the estimate of $\kappa$ is affected
by finite size effects, related to insufficient sampling of low-frequency acoustic modes: at 25 K these effects are so important, as to make the estimated
conductivity almost vanish (see below) 
\cite{footnote}.
%\footnote{We see a significant improvement in the estimate of $\kappa$ at 50 K from the 1728-atom model ($\kappa=0.027$ Wm$^{-1}$K$^{-1}$) to the 13824 model ($\kappa=0.25$ Wm$^{-1}$K$^{-1}$ see figure).
%However, at 50 K and lower temperatures the latter is not converged yet.}.
In fact, in order to eliminate finite-size effects, in our approach it would be necessary that in any relevant frequency range the density of vibrational states is larger than the normal-mode lifetimes, so that as many quasi-discrete normal modes as possible fall withing a line-width. In the low-frequency region, which is the most populated one in the quantum regime, this condition is hindered by the vanishing of both the density of states per unit volume and normal-mode line-widths. This effect is showcased in Figure~\ref{fig:tmp}, where we compare for different temperatures and model sizes the frequency-resolved differential conductivity,
\begin{equation}
    \kappa'(\omega) = \frac{1}{3V} \sum_\alpha\sum_{nm} \Delta(\omega-\omega_n) c_{nm} ( v_{n m}^{\alpha} )^2\tau^\circ_{nm}, \label{eq:kappa'(omega)}
\end{equation}
where $\Delta(\omega)$ is a broadened approximation of the $\delta$ function and the other symbols have the same meaning as in Eq. \eqref{eq:quantum}, and the conductivity accumulation function defined as:
\begin{equation}
    \kappa(\omega)=\int_0^\omega \kappa'(\omega')d\omega'. \label{eq:kappa(omega)}
\end{equation}

The AF model can also reproduce $\kappa$ for a-Si, but it is extremely sensitive to the empirical choice of the line broadening parameter ($\eta$). The impact of $\eta$ on the resulting $\kappa (T)$  is also shown in Figure~\ref{fig:kappa_quantum}, which shows that the value of $\kappa_{AF}$ varies by a factor two by varying $\eta$ between 0.01 and 0.5 meV in the temperature range considered. Whatever value is chosen for $\eta$, the AF model cannot reproduce the correct $\kappa_{QM}(T)$ of a-Si over the whole temperature range, in which we deem QHGK  accurate ($T\leq 600 K$), and it cannot give the correct decreasing trend at high temperature by construction. The predictions of the QHGK for the thermal conductivity of a-Si in the classical and fully quantum-mechanical regimes are compared in Figure \ref{fig:1728_classical_quantum} of SM.

In conclusion, we have introduced a unified approach to compute the lattice thermal con ductivity of both amorphous and crystalline systems. This quasi harmonic approach connects in a seamless fashion  the AF model for disordered systems and the BTE-RTA for crystals. QHGK provides a significant improvement in generality over the Allen-Feldman model for disordered systems and is analytically proven to be equivalent to BTE for periodic systems. Classical QHGK calculations were validated against MD simulations for a-Si, and yield satisfactory agreement over a wide temperature range. Quantum QHGK can be deemed predictive at low temperature, not only for glasses and crystals but also for partially disordered systems, for which parameter-free models were up to now unavailable. The technique proposed in this work paves the way to robust calculations of heat transport in systems with any kind of structural order, including materials with point defects, extended defects and nanostructuring, without relying on any implicit knowledge  of either their underlying symmetry, or the character of the vibrational modes, and without empirical parameters.

\begin{acknowledgments}
We thank Zheyong Fan for providing the GPUMD code and helping to set up the MD simulations. This work was partially funded by the EU through the \textsc{MaX} Centre of Excellence for supercomputing applications (Projects No.~676598 and 824143). While this paper was being written we learnt that conclusions similar to ours were reached by Simoncelli {\sl et al.}, following a different approach based on a generalization of the BTE \cite{Simoncelli:2019wy}.
\end{acknowledgments}

\putbib
\end{bibunit}

\onecolumngrid
\pagebreak
\clearpage

\setcounter{equation}{0}
\setcounter{figure}{0}
\setcounter{table}{0}
\setcounter{page}{1}

\renewcommand{\thesection}{S\arabic{section}}
\renewcommand{\thetable}{S\arabic{table}}
\renewcommand{\thefigure}{S\arabic{figure}}
\renewcommand{\theequation}{S\arabic{equation}}

\begin{bibunit}

\begin{center}
\textbf{\large Supplemental Material to \\
``Modeling heat transport in crystals and glasses \\ from a unified lattice-dynamical approach''}\\
\vspace{3.5mm}
Leyla Isaeva,$^1$~ Giuseppe Barbalinardo,$^2$ Davide Donadio,$^2$ and Stefano Baroni$^{1,3}$\\
\vspace{1.5mm}
\textit{$^1$SISSA -- Scuola  Internazionale  Superiore  di  Studi  Avanzati,  Trieste, Italy}\\
\vspace{0.5mm}
\textit{$^2$Department of Chemistry, University of California at Davis, USA}\\
\vspace{0.5mm}
\textit{$^3$CNR-IOM DEMOCRITOS, SISSA, Trieste, Italy}
\end{center}

\section{S1 -- Thermal conductivity in the classical  QHGK}
In order to establish Eqs. (3-5), we start from the expression for the harmonic heat flux, $J_\alpha$, Eq. (6), and Hamiltonian, $H$, in terms of the normal-mode complex amplitudes defined in the text:
\begin{equation}
\begin{split}
    \xi_n =\sum_{i\alpha} \sqrt{M_i}u_{i\alpha} e^\alpha_{ni}; \qquad
    \pi_n =\sum_{i\alpha} \frac{1}{\sqrt{M_i}}\dot{u}_{i\alpha} e^\alpha_{ni}; \qquad
    \alpha_n =\sqrt{\frac{\omega_n}{2}}\xi_n+\frac{i}{\sqrt{2\omega_n}}\pi_n; \\
    H = \frac{1}{2}\sum_n \left ( \pi_n^2 + \omega^2_n \xi_n^2 \right ) = \sum_n\omega_n |\alpha_n|^2; \qquad
    J =\frac{i}{2}\sum_{nm}v_{nm}\omega_m(\alpha^*_n+\alpha_n) (\alpha^*_m-\alpha_m),
    \label{seq:defs}
\end{split}
\end{equation}
where the Cartesian index of the flux in the second line of Eq. \eqref{seq:defs} has been overlooked not to mess with the notation of the complex amplitudes. The time evolution of the $\alpha$ amplitudes is:
\begin{equation}
  \alpha_n(t) = \alpha_n \mathrm{e}^{i\omega_nt},
\end{equation}
where $\alpha_n=\alpha_n(0)$ is the initial condition.
The product of the two fluxes appearing in Eq. (1) is a fourth-order polynomial in the $\alpha$'s and $\alpha^*$'s with time-dependent coefficients:
\begin{equation}
    J(t)J(0) = P_4(\alpha,\alpha^*;t)= -\frac{1}{4} \sum_{mnpq} v_{nm} \omega_m v_{pq} \omega_q  (\alpha^*_n(t)+\alpha_n(t)) (\alpha^*_m(t)-\alpha_m(t)) (\alpha^*_p + \alpha_p) (\alpha^*_q-\alpha_q).
\label{eq:flux_product}
\end{equation}
The canonical average of the above polynomial with respect to the harmonic Hamiltonian in Eqs. \eqref{seq:defs} is a Gaussian integral that can be evaluated using the Wick's theorem \cite{*[{See for instance Sec. 2.3 in }] [{}] Negele1988}, stating that the canonical average of a fourth-order monomial is equal to the sum of all the possible contractions:
\begin{equation}
    \langle ABCD \rangle = \langle AB\rangle \langle CD \rangle + \langle AC\rangle \langle BD \rangle + \langle AD\rangle \langle BC \rangle,
\end{equation}
where any of the capital letters above indicate any complex amplitude, $\alpha_n$ or $\alpha_n^*$. The relevant contractions are:
\begin{equation}
    \langle\alpha_n\alpha_m \rangle = 0; \qquad \langle \alpha_n^*(t)\alpha_m \rangle =\delta_{mn} g_n(t),
\end{equation}
where we define a single-mode classical Green's function $g_{n}(t) = \frac{k_BT}{\omega_n} e^{i\omega_n t}$. Hence, out of 16 terms in Eq. \eqref{eq:flux_product} only 6  are non vanishing. One such non-vanishing term is $\langle\alpha_n(t) \alpha_m(t) \alpha^*_p\alpha^*_q \rangle$, and the others are obtained by keeping two of the complex amplitudes conjugated. Making use of Wick's theorem this fourth-order correlator is reduced to the sum of the two terms:
\begin{equation}
    \langle\alpha_n(t) \alpha_m(t) \alpha^*_p \alpha^*_q \rangle = \langle \alpha_n(t) \alpha^*_p \rangle \langle \alpha_m(t) \alpha^*_q \rangle + \langle \alpha_n(t) \alpha^*_q \rangle \langle \alpha_m(t) \alpha^*_p \rangle  = (\delta_{np} \delta_{mq}  + \delta_{nq} \delta_{mp}) g_n(t) g_m(t).
\end{equation}
Working out the rest of the canonical averages in Eq. \eqref{eq:flux_product}, we may obtain the following relation for the left-hand side of the Eq. (S3) in terms of single-mode classical Green's functions $g_n(t)$ and $g_m(t)$:
\begin{equation}
\langle J(t)J(0) \rangle = -\frac{1}{4} \sum\limits_{nm}\left (
 \frac{\omega_n-\omega_m}{\omega_n}
 (g_{n}(t) g_{m}(t) +  g^*_{n}(t) g^*_{m}(t))
-\frac{\omega_n+\omega_m}{\omega_n}
 (g_{n}(t) g^*_{m}(t) + g^*_{n}(t) g_{m}(t)) \right).
\end{equation}
Introducing anharmonicity into our quasi-harmonic treatment through the linewidths, $\gamma_n$, of the vibrational normal modes results in the decay of the single-mode Green's functions $g_{n}(t)$ as:
\begin{equation}
    g_{n}(t) = \frac{k_BT}{\omega_n} e^{i(\omega_n + i \gamma_n)t}.
\end{equation}
By performing the time integrations and symmetrizing the final results, one obtains the heat conductivity tensor, represented in term of matrices ${v}_{nm}$ as
\begin{equation}
    \kappa_{\alpha\beta} = \frac{k_B}{V} \sum\limits_{nm} {v}^\alpha_{nm} {v}^\beta_{nm} {\tau}_{nm}, \qquad
    v^\alpha_{nm} = \frac{1}{2 \sqrt{\omega_{n} \omega_{m} }} \sum_{ij\beta\gamma} \frac{ R^{\circ}_{i\alpha} - R^{\circ}_{j\alpha}}{\sqrt{M_i M_j}} \Phi_{i\beta}^{j\gamma} e_n^{i\beta} e_m^{j\gamma}, \label{eq:sup_vnm}
\end{equation}
and matrix $\tau_{nm}$ given by the sum of two Lorentzian functions:
\begin{equation}
    \tau_{nm} =  \frac{(\omega_n+\omega_m)^2}{4 \omega_n \omega_m} \frac{\gamma_n+\gamma_m}{(\gamma_n+\gamma_m)^2 + (\omega_n-\omega_m)^2 } \\ + \frac{(\omega_n-\omega_m)^2}{4 \omega_n \omega_m} \frac{\gamma_n+\gamma_m}{(\gamma_n+\gamma_m)^2 + (\omega_n+\omega_m)^2 }. \label{eq:tau-complete}
\end{equation}
In the quasi-harmonic regime, linewidths are much smaller than normal-mode frequencies: $\epsilon=\frac{\gamma}{\omega}\ll 1$. In this regime, the second ``\emph{antiresonant}''term in Eq. \eqref{eq:tau-complete} can be neglected with respect to the first. To the same order in $\epsilon$, one has: $\frac{(\omega_n+\omega_m)^2}{4 \omega_n \omega_m} \approx 1 + \left ( \frac{\omega_n-\omega_m}{\omega_n+\omega_m}\right )^2$. By substituting this expression into Eq. \eqref{eq:tau-complete}, one gets: $\tau_{nm}=\tau^\circ_{nm}+\mathcal{O}\left ( \epsilon^2\right ) $, \emph{cfr.} Eq. (5).

\section{S2 -- Thermal conductivity in the quantum QHGK}
The derivation of the quantum QHGK expression for the heat conductivity follows the same path as in the classical case. To complete this derivation, we first introduce the quantum propagators $G_n(t)$ and $\tilde{G}_n(t)$ by promoting the classical complex amplitudes $\alpha_n,\alpha^*_m$ to the quantum ladder operators $\alpha_n \to \sqrt{\hbar} a_n, \alpha^*_m \to  \sqrt{\hbar} a^{\dagger}_m$ satisfying the Bose-Einstein commutation rule $[a_n, a^{\dagger}_m] = \delta_{nm}$:
\begin{equation}
    G_{n}(t) = \hbar \langle a^{\dagger}_n (t) a_n(0) \rangle
    =  \hbar n_k e^{i\omega_n t} ,\\ \qquad
    \tilde{G}_{n}(t) = \hbar \langle a_n (t) a^{\dagger}_n(0) \rangle
    =  \hbar (n_k+1) e^{-i\omega_n t}.
\end{equation}
We note that in the high-temperature limit the quantum single-mode Green's functions reduce to the classical one $\lim_{\hbar\to0} G_n(t)=\lim_{\hbar\to0} \tilde{G}^*_n(t)=g_n(t)$.
Next, in analogy with the classical case, we write the quantum canonical average $\langle \hat{J}(t)\hat{J}(0) \rangle$:
\begin{equation}
    \langle \hat{J}(\tau)\hat{J}(0) \rangle = -\frac{1}{4} \sum\limits_{nm}\left (  \frac{\omega_n-\omega_m}{\omega_n} \Bigl (G_{n}(\tau) G_{m}(\tau) +  \tilde{G}_{n}(\tau) \tilde{G}_{m}(\tau) \Bigr ) -\frac{\omega_n+\omega_m}{\omega_n} \Bigl (G_{n}(\tau) \tilde{G}_{m}(\tau) + \tilde{G}_{n}(\tau) G_{m}(\tau) \Bigr ) \right),
\end{equation}
where $\tau=t+i\hbar\lambda$ is the complex argument of the quantum GK formula, Eq. (7). By introducing finite mode linewidths and performing the double time integration in Eq. (7), we arrive at the following
lengthy relation for the thermal conductivity tensor:
\begin{multline}
 \kappa_{\alpha\beta} = \frac{\hbar^2}{4VT}\sum\limits_{nm} v^{\alpha}_{nm} v^{\beta}_{nm} \left ( \frac{ n_n - n_m } {\hbar(\omega_n-\omega_m)}    \frac{\gamma_n+\gamma_m}{(\omega_n-\omega_m)^2+(\gamma_n+\gamma_m)^2}       (\omega_n+\omega_n)^2  \right . \\ \left . + \frac{e^{\beta\hbar(\omega_n+\omega_m)}-1} {\hbar(\omega_n+\omega_m)} \frac{2 (\omega_m-\omega_m)^2}{(e^{\beta\hbar\omega_n}-1)(e^{\beta\hbar\omega_m}-1)}
        \frac{\gamma_n+\gamma_m}{(\omega_n+\omega_m)^2+(\gamma_n+\gamma_m)^2}
         \right ) \label{seq:quantum-QHGK}
\end{multline}
The second, antiresonant, term in Eq. \eqref{seq:quantum-QHGK} can be neglected in the quasi-harmonic regime $\bigl ( \frac{\gamma}{\omega}\to 0 \bigr )$, while the first one can be cast into a BTE-like form by introducing the matrix $c_{nm} = \frac{\hbar\omega_n\omega_m}{T} \frac{n_n-n_m}{\omega_n-\omega_m}$:
\begin{equation}
   \kappa_{\alpha \beta} = \frac{1}{V} \sum_{nm} c_{nm} v_{n m}^{\alpha} v_{nm}^{\beta} \tau^\circ_{n m}. \label{seq:quantum}
\end{equation}

\section{S3 -- Computational details}

\begin{figure}[!tbp]
\includegraphics[width = 0.49\textwidth]{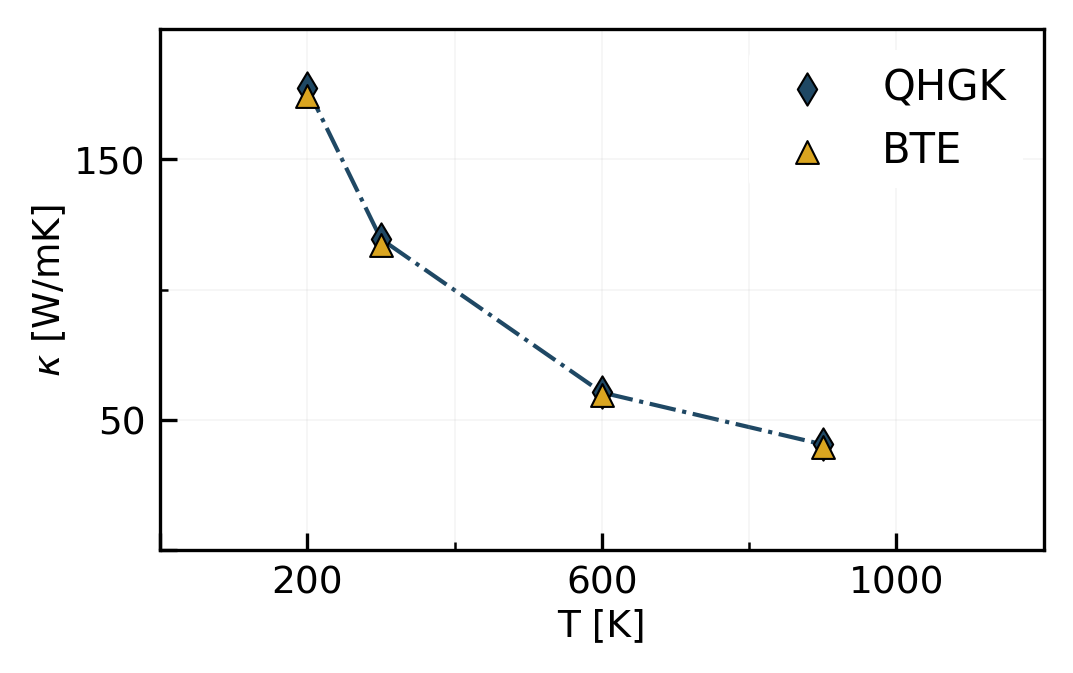}
\caption{Thermal conductivity of $fcc$ Si computed for a 1728-atom supercell using our QHGK approach (blue) in comparison with the standard BTE calculations (yellow).}
\label{fig:gk_bte_comparison}
\end{figure}

\begin{figure}[!tbp]
\centering
% \begin{subfigure}{0.45\textwidth}
%   \centering
%   \includegraphics[width =\linewidth]{FigS2a.png}
%     \caption{ }
%     \label{fig:linewidths}
% \end{subfigure}\hspace{7mm}
% \begin{subfigure}{0.34\textwidth}
%   \centering
%   \includegraphics[width=\linewidth]{FigS2b.png}
%     \caption{ }
%     \label{fig:heatmap_tau}
% \end{subfigure}
%\setbox1=\hbox to 0.45\textwidth{\vsize=\ht0 \vfill \includegraphics[width=\hsize]{FigS2b.png} \vfill}
\setbox1=\hbox{\includegraphics[width =0.45\linewidth]{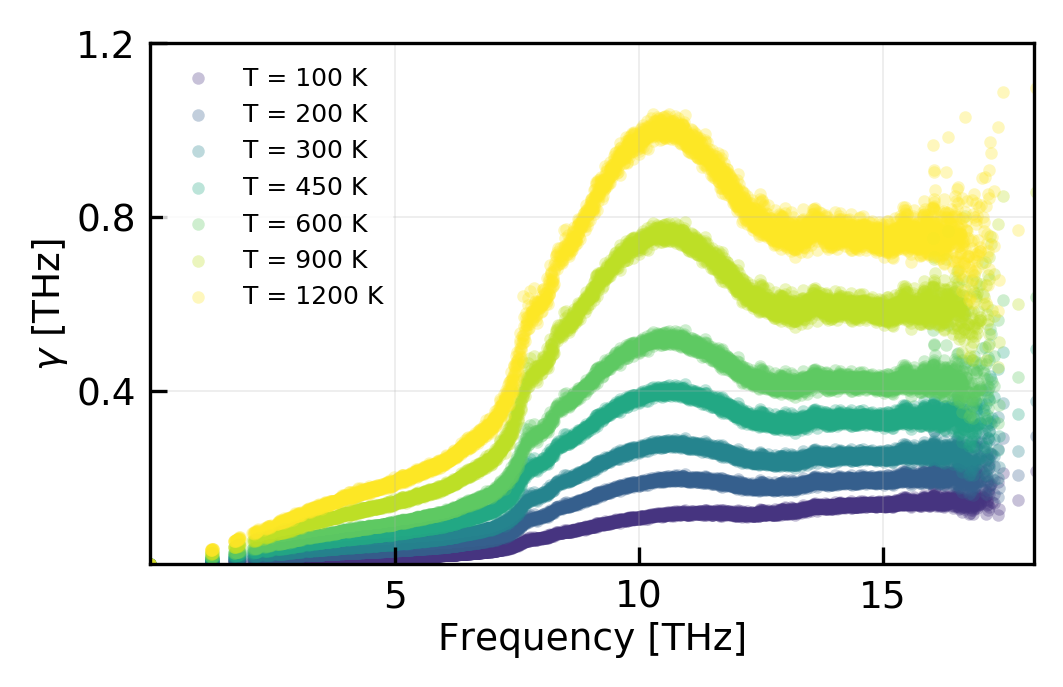}}
\setbox2=\hbox{\includegraphics[width =0.4\linewidth]{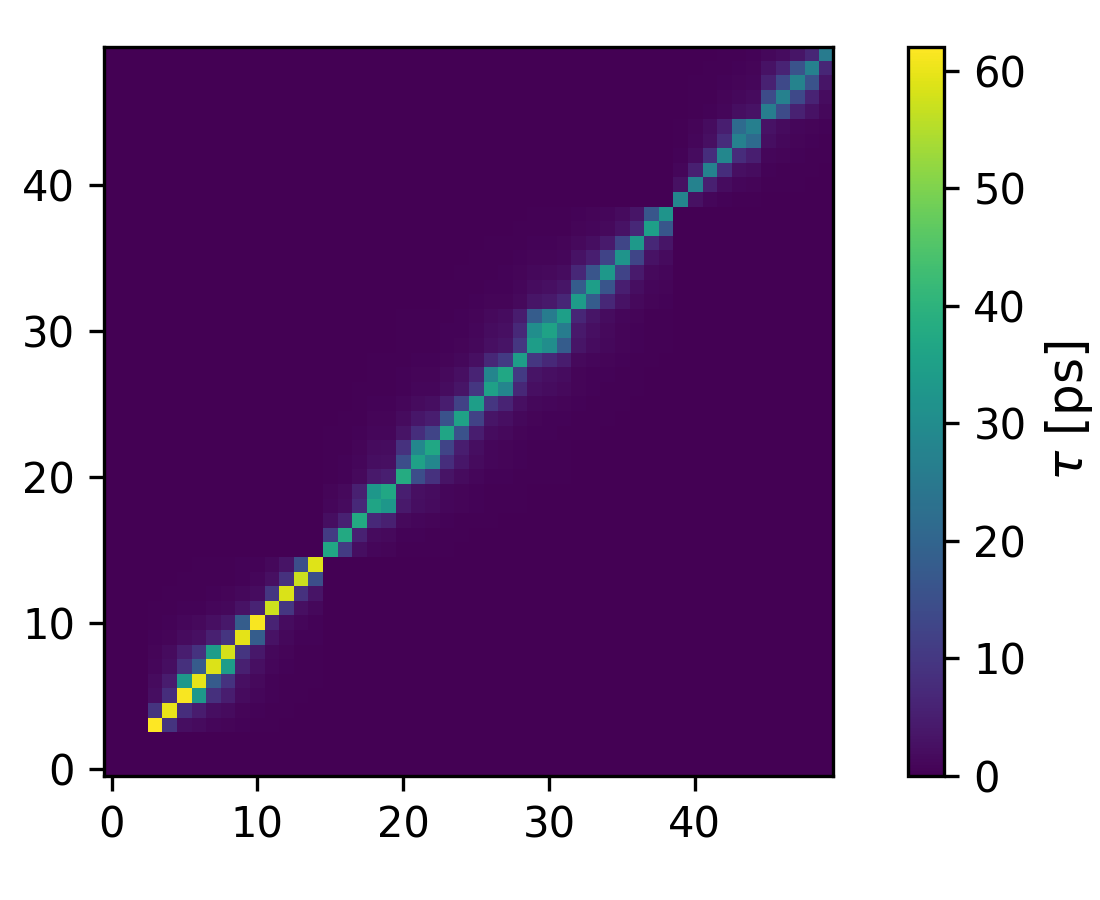}}
\hbox to \textwidth{\hfill\vbox to \ht2{\vfill\box1\vfill}\hfill\box2}
\caption{ Left: Normal-mode linewidths of 1728-atomic model of a-Si
computed for various temperatures using Fermi golden rule
and the classical limit of the Bose-Einstein occupation function, which corresponds to equipartition. Right: Partial heatmap of matrix $\tau_{nm}$ computed for $T$=300 K. The structure of the matrix, \emph{i.e.} non-vanishing diagonal and close-to-diagonal elements, is dictated by its analytical form given by the Lorentzian function.}
\label{fig:linewidths}
\end{figure}

\begin{figure}[!tbp]
\centering
\includegraphics[width=0.5\textwidth]{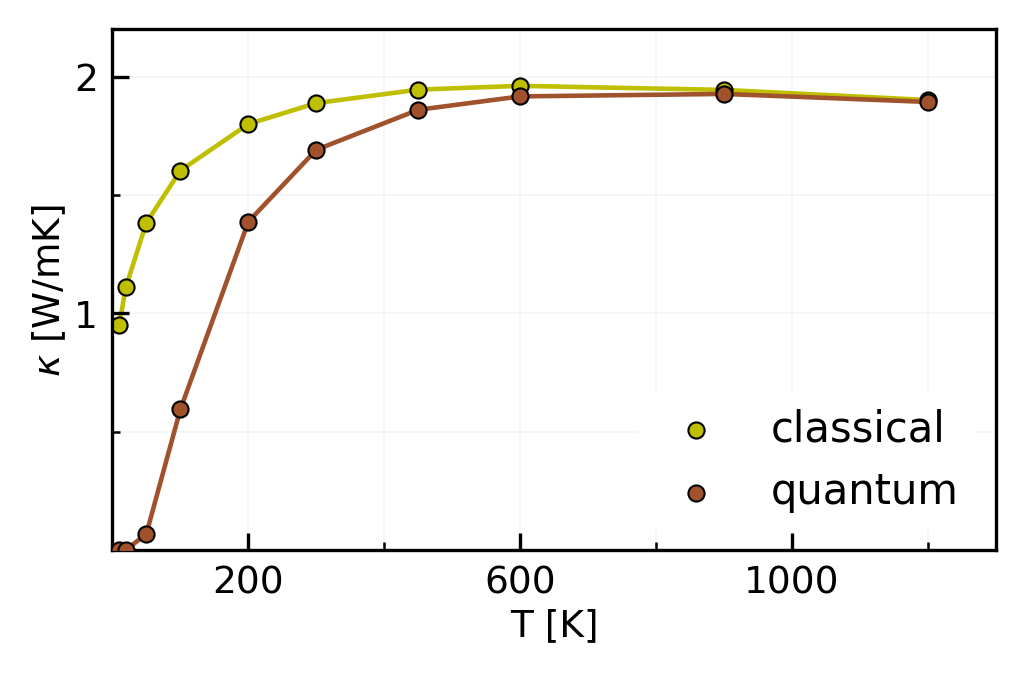}
\caption{Thermal conductivity computed for 1728-atomic model
of a-Si in classical and quantum regimes.}
\label{fig:1728_classical_quantum}
\end{figure}

\begin{figure}[!tbp]
\includegraphics[width = 0.49\textwidth]{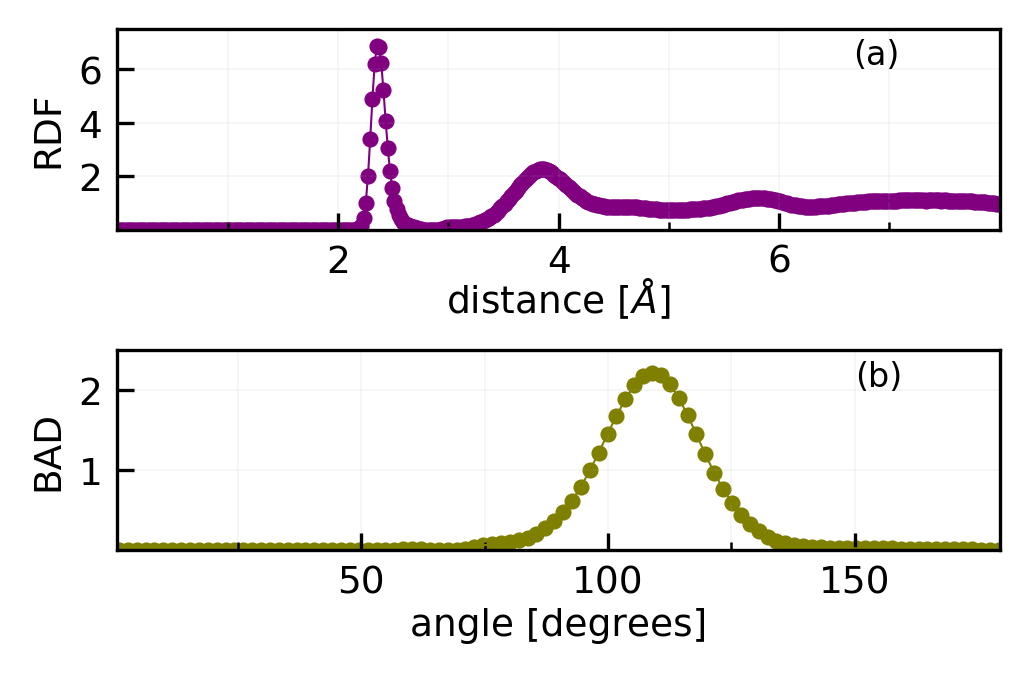}
\caption{(a) Radial distribution function (RDF) and (b) bond angle distribution (BAD) of the 1728 atoms model of a-Si used for the calculations of thermal conductivity.}
\label{fig:gofr}
\end{figure}

A liquid model prepared at 3000 K is quenched and then equilibrated to 2000 K for 10 ns at constant pressure. It is then further quenched to 300 K at constant volume in 10 ns, and equilibrated at the same temperature for 1 ns. This procedure produces a-Si models of good quality with a very low concentration of coordination defects~\cite{Deringer:2018in}. The model is a cubic simulation box with a density of 2.3 g/cm$^3$. {The resulting radial and bond-angle distribution functions are reported in Figure \ref{fig:gofr}.} The average coordination is 4.06 neighbours per atom, indicating that the system can be considered as a random tetrahedral network.

For this model we calculate $\kappa$ at several temperatures between 100 K and 1200 K by equilibrium MD simulations implementing Eq.(1) according to GK theory. Starting from the model at 300 K, the system is equilibrated at the target temperatures for 1 ns at fixed volume before each production run. The latter is carried out integrating the equations of motion in the microcanonical ensemble (NVE) with a timestep of 0.5 fs for a total of 25 ns.
All MD simulations are performed using the GPUMD open-source code~\cite{Fan2017}, calculating the heat flux $J$ every 4 fs~\cite{Fan:2015ba}.

The thermal conductivity was extracted from the energy flux thus generated, using the recently introduced \emph{cepstral analysis} method \cite{Ercole2017,Bertossa2019}. Cepstral analysis \cite{Bogert1963,*Childers1977} is a technique, commonly used in signal analysis and speech recognition, to process the power spectrum of a time series, leveraging its smoothness and the statistical properties of its samples. According to Eq.~(1) of the main text, the thermal conductivity is proportional to the zero-frequency value of the power spectrum of the energy flux: $\kappa\propto S(\omega=0)$, where $S(\omega)=\int_{-\infty}^\infty \mathrm{e}^{i\omega t} C(t)dt$, and $C(t) = \langle J(t) J(0)\rangle$ is the flux time auto-correlation function. The Wiener-Kintchnine theorem \cite{Wiener1930,*Khintchine1934} states that $ S(\omega)$ is asymptotically proportional to the expectation of the squared modulus of the truncated Fourier transform of the flux sample: $ S(\omega) = \lim_{\tau\to\infty} \langle \frac{1}{\tau}  | \tilde {\mathcal J}_\tau(\omega)|^2 \rangle$, where $\tilde{J}_\tau(\omega) = \int_0^\tau J(t) \mathrm{e}^{i\omega t}dt$. In the long-time limit, the squared modulus to be averaged is a stochastic process whose values are independent for $\omega\ne\omega'$ and individually distributed as $\frac{1}{\tau}  | \tilde { J}_\tau(\omega)|^2 = S(\omega) \xi(\omega) $, where $\xi(\omega) \sim \frac{1}{2}\chi^2_2$, $\chi^2_2$ being a chi-square variate with two degrees of freedom. The multiplicative nature of the noise affecting the sample spectrum suggests that the power of the noise can be reduced by applying a low-pass filter to its logarithm. This is the main idea underlying cepstral analysis, which can be leveraged to devise a consistent and asymptotically unbiased estimator for the the zero-frequency value of the flux power spectrum, which is proportional to the transport coefficient we are after. For more details, see Refs. \onlinecite{Ercole2017}, \onlinecite{Baroni2018}, and \onlinecite{Bertossa2019}.
% The cutoff frequency of the low-pass filter was set to 1.5 THz and the width of the window function was chosen as 0.1.
Given the strongly harmonic nature of the system at low and intermediate temperatures, in order to improve the sampling of the phase space at 300 K and below, we average  the results obtained by cepstral analysis over two independent simulations 25 ns long.

In order to implement the classical QHGK approach as in Eq.(4), we optimize the a-Si model structure by steepest descent and calculate the second- and third-order force constant matrices by finite differences (frozen phonon method) with atoms displacements of $10^{-4}$ \AA. Normal modes line widths $\gamma_n$, necessary to evaluate Equation 6 for $\tau_{nm}$, are computed using the Fermi golden rule \cite{Fabian1996} (See Figure \ref{fig:linewidths}). In the disordered case, this is done explicitly only for the smaller (1728-atom) sample. For larger samples, we interpolate the inverse linewidth, i.e. lifetimes, as a function of frequency from the explicit results for the 1728 atoms system. Lifetimes vs. frequencies are averaged over frequency bins and then interpolated with third-order splines, with the constraint that at low frequency $\tau \propto 1/\omega^2$ \cite{AsenPalmer:1997bl}. When comparing with classical MD simulations, QHGK results were obtained using the classical limit of the normal-mode lifetimes; the full quantum expression was used otherwise.

\putbib
\end{bibunit}


\begin{thebibliography}{31}
\bibitem [{foo()}]{footnote2}%
  \BibitemOpen
  \bibinfo {note} {The temperature
  dependence in the AF model lies only in the heat capacity term, therefore in
  the classical limit $\kappa$ is temperature independent.}\BibitemShut {Stop}%
\bibitem [{foo()}]{footnote}%
  \BibitemOpen
  \bibinfo {note} {We see a significant
  improvement in the estimate of $\kappa$ at 50 K from the 1728-atom model
  ($\kappa=0.027$ Wm$^{-1}$K$^{-1}$) to the 13824 model ($\kappa=0.25$
  Wm$^{-1}$K$^{-1}$ see figure). However, at 50 K and lower temperatures the
  latter is not converged yet.}\BibitemShut {Stop}%
\end{thebibliography}
\end{document}